\documentclass[showpacs,aps,pra,twocolumn,amsmath,amssymb,superscriptaddress]{revtex4-1}
\usepackage{amsmath}           % Package
\usepackage[latin1]{inputenc}  % Accents
\usepackage[dvipdf]{epsfig}    % Encapsulate Postscript

\usepackage{doipubmed}

\begin{document}

\title{Green function approach for scattering quantum walks}

\author{F. M. Andrade}
\email{fmandrade@uepg.br}
\affiliation{Departamento de Matem\'atica e Estat\'{\i}stica,
Universidade Estadual de Ponta Grossa,
84030-900 Ponta Grossa-PR, Brazil}
\author{M. G. E. da Luz}
\email{luz@fisica.ufpr.br}
\affiliation{Departamento de F\'{\i}sica,
Universidade Federal do Paran\'a,
C.P. 19044, 81531-980 Curitiba-PR, Brazil}

\date{\today}

\begin{abstract}
In this work a Green function approach for scattering quantum walks 
is developed. 
The exact formula has the form of a sum over paths and always
can be cast into a closed analytic expression for arbitrary topologies 
and position dependent quantum amplitudes.
By introducing the step and path operators, it is shown how to extract
any information about the system from the Green function.
The method relevant features are demonstrated by discussing in 
details an example, a general diamond-shaped graph.
\end{abstract}



\maketitle

\section{Introduction}

Generally speaking, quantum walks (QW) represent unitary evolutions 
taking place in discrete spaces -- graphs -- for which typical basis
states are localized. 
There are several ways to formulate QW, either considering time as
a continuous (CTQW)  \cite{farhi,childs-time} or a discrete variable. 
In the latter case, the two major formulations are the (a)
coined QW (CQW), based on inner ``coin'' states (see, e.g., \cite{tregenna}), 
and (b) scattering QW (SQW), relying on the idea of multi-port 
interferometers \cite{hillery-pra, feldman-jpa}.
The continuous time and coined QW are directly related as a limit
process \cite{relation-coin-continuous}, whereas the CQW and SQW 
have been shown to be unitarily equivalent in arbitrary topologies 
\cite{andrade}.

Quantum walks originally emerged \cite{aharonov-meyer-watrous} 
from the interest to construct and understand quantum analogs 
of classical random walks (CW). 
But soon it was realized they also would constitute powerful tools 
in quantum computation \cite{proceedingsacm}, specially given that
QW can represent universal quantum computation primitives 
\cite{childs-lovett}.
In fact, for a long time CW have been used to solve different 
computational problems \cite{motwani}.
Thus, the connections between the quantum and classical walks 
\cite{mackay,konno1,kosik}, allied to the particular features of 
the former \cite{kempe}, actually point to the potential usefulness 
of QW in building algorithms which are much faster and robuster 
\cite{mosca,ambainis-1} than their classical counterparts.
As representative examples we can cite the Grover algorithm 
\cite{grover} (simulated through QW \cite{shenvi}) for searching 
of unsorted database, the element distinctness algorithm \cite{ambais}, 
the detection of marked elements \cite{szegedy-magniez}, the 
computation of orders of solvable groups \cite{watrous-groups},
and the quantum Fourier transform \cite{hines}.
Moreover, even problems like the energy transport in biological
systems can be analyzed by means of QW \cite{biology}.

A key aspect in such class of systems is the quantum interference 
between the possible ``paths'' (see next Section) along the evolution 
\cite{knight,kendon,lee,oka}. 
It leads to a dynamics that generally spreads much faster than 
CW \cite{kempe} (although in certain situations anomalous 
sub-diffusive behavior may also emerge \cite{prokof}).
As a consequence, one gets exponentially faster hitting times
from QW \cite{farhi,childs,kempe-hitting}, one of the reasons 
why QW are particularly suitable \cite{gabris-reitzner} to solve 
searching problems \cite{hillery,lee}. 
Also, different diffusion processes, from ballistic to Anderson 
localization \cite{schreiber,ahlbrecht}, are possible when 
decoherence is included.

Since interference is fundamental to explain different
phenomena observed in QW \cite{kendon-philosophical} (including 
many of the applications mentioned above) it is desirable to have
a description emphasizing the path-like character of QW. 
In this respect Green function methods are particularly 
useful \cite{luz1,schmidt1,schmidt2}.
Then, here we develop a full Green function approach for QW in 
arbitrary topology and for position dependent quantum amplitudes.
For so, we assume the very appropriate discrete scattering 
formulation, SQW.
We should observe there are few interesting works (e.g.,
Refs. \cite{carteret,konno2,konno3}) addressing the classification
of trajectories in QW.
They, nevertheless, are based mostly on combinatorial analysis to 
compute all the possible final states at a time $t = n$.
Our proposed construction thus is much closer to the idea of 
``history'' of trajectories in the Feynman sense \cite{feynman}.

The paper is organized as the following.
In Sec. II we review the scattering formulation for quantum
walks, also making few useful parallels with the classical case.
By direct mapping one-dimensional QW to a related type of problem,
1D point interaction lattices, in Sec. III we are able to write the 
exact Green function $G$ in the form of a sum over paths.
Moreover, we discuss how such formula can be summed as a closed 
analytical expression.
In Sec. IV the 1D construction is extended to complete arbitrary
topologies.
By defining the step and path operators, we show in Sec. V how to 
extract any system relevant information from the exact expression 
for $G$.
In Sec. VI we illustrate the features of the present approach  
analyzing in details a particular example, the diamond-shaped 
graph.
Finally, we present the conclusion in Sec. VII.

\section{A brief review on the scattering formulation for quantum walks}

To better understand the main ideas underlying the definition of
quantum walk models, and thus to develop a Green function approach, 
here we review QW scattering formulation \cite{hillery-pra} 
on the line (1D).
The case of more general topologies will be discussed in
the next Sections.

So, consider a helpful framework for SQW: to view their evolution as 
a dynamics defined on a 1D ``Hilbert lattice'', depicted in Fig. 1.
Notice, however, it does not necessarily represent a spatial structure 
since the states (assumed on the bonds) do not need to be position 
eigenvectors.
Under this picture, the lattice characteristic parameter is 
$L = \Delta{j} = 1$, just the spacing between two consecutive sites 
of the Hilbert lattice.
Along each bond, joining the sites $j$ and $j+1$ (Fig. 1), we have 
two possible states, $|+1,j+1 \rangle$ and $|-1,j \rangle$.
Then, each basis element, $|\sigma,j \rangle$, is labeled by two 
quantum numbers.
The first, $\sigma$, sets the ``direction'' ($\pm 1$) along the 
lattice.
We mention that although fully equivalent, the present is slight 
different than the common SQW construction in the literature.

\begin{figure}
\centerline{\psfig{figure=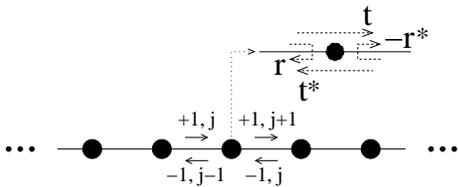,width=6cm}}
\caption{The ``Hilbert lattice'' associate to SQW in 1D.
For each site it is defined appropriate scattering 
quantum amplitudes (here illustrating the phase convention in 
\cite{hillery-pra}).}
  \label{fig1}
\end{figure}
\medskip

The discrete time evolution is given by the one step
unitary operator $U$, such that
$| \Psi(n+1) \rangle = U | \Psi(n) \rangle$.
For $U$, we consider the translation ($S_T$) and 
reversion-translation $(S_{RT})$ operators
\begin{eqnarray}
S_T |\sigma,j \rangle &=&
|\sigma, j + \sigma \rangle, \ \ \
{S_T}^{\dagger} |\sigma,j \rangle =
|\sigma,j -\sigma \rangle,
\nonumber \\
S_{RT} |\sigma,j \rangle &=& {S_{RT}}^{\dagger} |\sigma,j \rangle
= |-\sigma, j - \sigma \rangle,
\label{translation-reflection}
\end{eqnarray}
with both unitary and ${S_{RT}}^2 = {\bf 1}$.
We also define $T$ and $R$, for which the basis states
$|\sigma,j \rangle$ are eigenstates, or
\begin{equation}
T |\sigma,j \rangle =
t_{\sigma, j} |\sigma,j \rangle, \qquad 
R |\sigma,j \rangle =
r_{\sigma, j} |\sigma,j \rangle.
\end{equation}
If we impose now 
\begin{eqnarray}
& &r_{+1, j} \, t_{+1, j}^{*} + r_{-1, j}^{*} \, t_{-1, j} = 
   r_{+1, j} \, t_{-1, j}^{*} + r_{-1, j}^{*} \, t_{+1, j} =
0, \nonumber \\
& & |t_{\pm, j}|^2 + |r_{\pm, j}|^2 = 
    |t_{\pm, j}|^2 + |r_{\mp, j}|^2 = 1, 
\label{condition-coef}
\end{eqnarray}
then, the unitary time evolution reads ($0 \leq \gamma < 2 \pi$)
\begin{equation}
U(\gamma) = \exp[i \gamma] \, \big( S_T \, T + S_{RT} \, R \big).
\label{evolution-qrw}
\end{equation}
The term $\exp[i \gamma]$ is associated to the translation between 
neighbor sites ($\Delta j = L = 1$), relevant to proper describe 
stationary scattering solutions \cite{feldman-pla}.

Provided Eq. (\ref{condition-coef}) holds, there is a freedom 
to choose the coefficients $r_j$ and $t_j$. 
For instance, by setting
($0 \leq \rho_j \leq 1$ and $0 \leq \phi_j, \varphi_j < 2 \pi$ 
for any $j$)
\begin{equation}
t_{\sigma, j} = \rho_j \exp[i \sigma \phi_j],
\qquad
r_{\sigma, j} = \sigma \sqrt{1-\rho_j^2}
\exp[i \sigma \varphi_j],
\label{eigenvalues}
\end{equation}
one gets $r_{-1, j} = -r_{+1, j}^{*}$ and  $t_{-1, j} = t_{+1, j}^{*}$, 
just the convention used in \cite{hillery-pra} (Fig. 1).

The dynamics in Eqs. (\ref{translation-reflection})--(\ref{evolution-qrw}) 
in fact represents an extended quantum version of a more simple classical 
random walk.
Each time the classical walk needs to choose 
a new direction, it uses the same probabilities ($P$ and $1-P$) 
to decide between right and left.
By allowing in Eq. (\ref{eigenvalues}) $\rho$ and the phases to 
depend on $j$, we are implicitly assuming position dependent 
distribution functions for the direction choices.
Obviously, by setting a same $\rho$, $\varphi$ and $\phi$
for any $j$ we recover the usual case.

Finally, as it stands the above model is deterministic in the quantum 
mechanical sense:
any initial state $|\Psi(0)\rangle$, after $n$ time steps,
is uniquely determined by the always well defined state 
$U^n |\Psi(0)\rangle$.
Thus, stochasticity (i.e., classical randomness) can enter into the 
problem only through measurements, when we determine the system
location along the Hilbert lattice.
In fact,
\begin{equation}
P_{\sigma, j}(n) = |\langle j, \sigma |\Psi(n)\rangle|^2
\end{equation}
is the probability to be in the quantum state (or in the present 
lattice language ``position and direction'') $j, \sigma$ at time $n$.
So, projection is an essential ingredient in QW.

As a simple example, consider the initial state 
$|\Psi(0)\rangle = |+1,0\rangle$.
Under $U$ one has after $n=3$ time steps
\begin{widetext}
\begin{eqnarray}
 |\Psi(3)\rangle &=& \exp[3 i \gamma] 
\Big\{
t_{+1,0} t_{+1,1} t_{+1,2} \, |+1,3\rangle +
(r_{+1,0} r_{-1,-1} t_{+1,0} + t_{+1,0} r_{+1,1} r_{-1,0})
|+1,1\rangle +
r_{+1,0} t_{-1,-1} r_{-1,-2} \, |+1,-1\rangle \nonumber \\
& & +
r_{+1,0} t_{-1,-1} t_{-1,-2} |-1,-3\rangle +
(r_{+1,0} r_{-1,-1} r_{+1,0} + t_{+1,0} r_{+1,1} t_{-1,0})
|-1,-1\rangle +
t_{+1,0} t_{+1,1} r_{+1,2} |-1,1\rangle
\Big\}.
\label{example-3steps}
\end{eqnarray}
\end{widetext}
Thus, the system probability to be found, say, in
$|+1,3\rangle$ is $|t_{+1,0} t_{+1,1} t_{+1,2}|^2$.
Note that for three time steps, there is only one possible ``path'' 
ending up in $|+1,3\rangle$.
Hence, the modulus square of the quantum amplitude associated to 
such path yields the sough probability. 
On the other hand, there are two possible paths leading
to $|-1,-1\rangle$.
They correspond to the amplitudes $r_{+1,0} r_{-1,-1} r_{+1,0}$ and 
$t_{+1,0} r_{+1,1} t_{-1,0}$ (cf. Eq. (\ref{example-3steps})).
But contrary to CW, where the total probability is the sum of the 
individual probabilities of each trajectory, here the quantum 
interference character of the walk demands that 
$P_{-1,-1}(3) = |r_{+1,0} r_{-1,-1} r_{+1,0} + t_{+1,0} r_{+1,1} t_{-1,0}|^2$.

%\squeezetable
%
\begin{table*}[t]
\label{tableI}
\caption{The correspondence between quantities 
in the complete biased quantum walk and in the 1D free propagation}
%\footnotesize{
\begin{tabular}{l l}
\hline
\hline
Fully biased 1D quantum walk \ \ \ \ \ \ \ \ \ \ \ \ &
Free quantum propagation on the line
\\ \hline
%LINE 1
$U = \exp[i \gamma] \, S_T$
&
${\mathcal U}{(\tau)} = \exp[-i (\hat{p}^2/2) \tau]$,
\ \ $\tau = L/v_{\mbox{\scriptsize phase}} = L/(p/2)$,
\ \ $p = k$,
\ \ $L = 1$
\\
%LINE 2
$ |\Psi(0)\rangle = \frac{1}{\sqrt{2 \pi}}
\sum_{j=-\infty}^{j=+\infty} \exp[i j \gamma]
|+1, j\rangle$
&
$|\Phi(0)\rangle = |p\rangle =
\frac{1}{\sqrt{2 \pi}} \int_{-\infty}^{+\infty}
dx \exp[i p x] |x\rangle$
\\
%LINE 3
$|\Psi(1)\rangle = U|\Psi(0)\rangle =
|\Psi(0)\rangle$
\footnote{%The correspondence between $U$ and ${\mathcal U}{(\tau)}$
%is given up to a phase. 
Strictly speaking, $\exp[i \gamma]$ in the definition of $U$ in Eq. 
(\ref{evolution-qrw}) is not necessary and then the equivalence  
would be complete (here, in each step it is up to such global phase).
Nevertheless, if we take as an initial state 
$|\Psi(0)\rangle = |\sigma, j\rangle$, 
instead of states like Eq. (\ref{initial}), $\exp[i \gamma]$
becomes useful to study interference in more general topologies, 
as in Sec. VI. 
The important point, therefore, is that we have just a phase difference, 
so not compromising any parallel between the systems time evolutions.}
&
$|\Phi(\tau)\rangle = 
{\mathcal U}(\tau) |\Phi(0)\rangle =
\exp[-i p^2 \tau/2]
|\Phi(0)\rangle = \exp[-i p] |\Phi(0)\rangle$ 
\\
%LINE 4
$\gamma$
&
$p$
\\
\hline \hline
\end{tabular}
%}
\end{table*}

\section{A Green function approach for quantum random walks}

\begin{figure}
\centering
\centerline{\psfig{figure=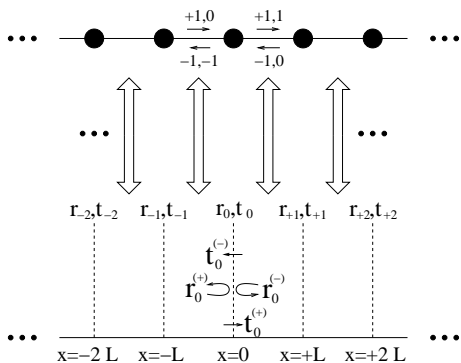,width=6cm}}
\caption{Schematic association between QW Hilbert (space) lattice 
and a generalized Kronig-Penney configuration lattice.
Each site $j$ corresponds to a point interaction at $x = j L$ 
(of reflection and transmission amplitudes $r_{j}^{(\pm)}$ and 
$t_{j}^{(\pm)}$).}
\label{fig2}
\end{figure}

Here we develop a Green function approach for the SQW in the 
previous Section, proceeding in three steps.
(a) First, we construct a mapping from our QW to a 1D
generalized Kronig-Penney lattice \cite{schmidt1}, for which we 
can calculate the exact energy-dependent Green function $G$.
(b) Then, we discuss which are the mapped system appropriate 
configurations in order to match the original problem.
(c) Finally, we show how the obtained $G$ gives the quantum walk
sought dynamics.
We leave to the next Section the extension of the 1D 
results to the case of more general topologies, namely, QW on 
arbitrary graph structures.

\subsection{The mapping}

As already emphasized, the quantum walk does not necessarily 
represent any dynamics on a concrete physical lattice.
Nevertheless, for our purposes it is very useful to associate the 
quantum walk Hilbert (space) lattice -- and its underlying 
``kinematics'' \cite{balazs} -- to that of an usual continuous 1D 
quantum scattering problem.

In Fig. 2 we show schematically the correspondence between the 
model of Fig. 1 with a generalized Kronig-Penney lattice of 
equally spaced arbitrary point interactions \cite{schmidt1}, 
i.e., zero range potentials which extend the usual delta function
\cite{albeverio}.
Each point interaction (at $x = \pm j L$, $j = 0, 1, \ldots$) 
is entirely characterized by the quantum amplitudes $r_j^{(\pm)}(k)$ 
and $t_j^{(\pm)}(k)$.
The superscript $+$ ($-$) stands for the reflection or transmission
of a plane wave of wave number $k$ incoming from the left (right) 
of the point interaction location.
Hereafter, subscripts (superscripts) for direction quantum numbers
indicate that the corresponding $r$'s and $t$'s
are those for QW (continuous scattering) systems.
For the most general zero range potential, we have that 
(see, e.g., Ref. \cite{zanetti} for a full discussion)
\begin{eqnarray}
r_j^{(\pm)}(k) &=& 
\frac{c_j \pm i k (d_j - a_j) + b_j k^2}{-c_j + i k (d_j + a_j) + b_j k^2}
\, \exp[\pm i k e_j],
\nonumber \\
t_j^{(\pm)}(k) &=& 
\frac{2 i k \exp[\pm i \theta_j]}{-c_j + i k (d_j + a_j) + b_j k^2},
\label{point-interaction}
\end{eqnarray}
where $a_j \, d_j - b_j \, c_j = 1$, with $a_j, b_j, c_j, d_j, e_j$ 
real and $\theta_j \in [0, 2 \pi)$ \cite{comment-coef}. 
Equation (\ref{point-interaction}) satisfies the relations in 
Eq. (\ref{condition-coef}) and also to
\begin{equation}
r_j^{(\pm)}(k) = {r_j^{(\pm)}}^*(-k), \qquad 
t_j^{(\pm)}(k) = {t_j^{(\mp)}}^*(-k).
\label{rr-tt}
\end{equation}
Furthermore, for $b_j = c_j = 0$, they become independent on
$k$ (up to the phases for the $r$'s) and Eq. (\ref{point-interaction}) 
assumes the same form as Eq. (\ref{eigenvalues}).

Now, let us set $m = \hbar = 1$ (so $p = k$), define 
$\tau = L/v_{\mbox{\scriptsize phase}}$ with $v_{\mbox{\scriptsize phase}} = 
p/2$, and for convenience take $L = 1$.
Then, we can make a direct association between the quantum walk
one step evolution operator $U$ and the continuous system propagator 
${\mathcal U}{(\tau)}$, 
mapping $U|\Psi(0)\rangle = |\Psi(1)\rangle$ to 
${\mathcal U}(\tau) |\Phi(0)\rangle = |\Phi(\tau)\rangle$.

To concretely establish the correspondence, we start with the 
simplest situation of a fully biased quantum walk, i.e., one
which always evolves to a same direction.
We thus assume $\rho_j = 1$ and $\phi_j = 0$ for any $j$, 
from Eq. (\ref{evolution-qrw}) leading to  
$U = \exp[i \gamma] \, S_T$.
Such case presents a close parallel with a quantum particle 
propagating freely along the line.
In our generalized Kronig-Penney lattice, a free particle is 
trivially obtained by setting all the reflection (transmission) 
amplitudes equal to 0 (1), so that the time evolution is   
${\mathcal U}{(t)} = \exp[-i (\hat{p}^2/2) t]$, with
$\hat{p} |p\rangle = p |p\rangle$ for $|p\rangle$ the moment 
eigenstate.
Hence, we have a direct mapping between the complete biased quantum 
walk dynamics and the evolution of a free particle on the line for 
$t = \tau$.
The equivalent quantities are listed in Table I.

Next, we consider that in Eq. (\ref{eigenvalues}) for any $j \neq 0$ 
we have $\rho_j = 1$ and $\phi_j = 0$, and for $j=0$ we have 
arbitrary $\rho$ and phases.
Also, we assume as the quantum walk initial state
\begin{equation}
|\Psi(0)\rangle = \frac{1}{\sqrt{2 \pi}} \sum_{j=-\infty}^{j=0}
\exp[i j \gamma] |+1, j\rangle,
\label{initial}
\end{equation}
so that $P_j(n = 0) = 0$ for $j > 0$.
Then, applying $n$ times the evolution operator, Eq.
(\ref{evolution-qrw}), to $|\Psi(0)\rangle$ we get 
(with $r = r_{+1,0}$ and $t = t_{+1,0}$)
\begin{widetext}
\begin{equation}
|\Psi(n)\rangle =
U^n |\Psi(0)\rangle =
\frac{1}{\sqrt{2 \pi}}  \Big\{
\sum_{j=-\infty}^{j=0} \exp[i j \gamma] |+1, j\rangle
+
r \sum_{j=-n}^{j=-1} \exp[-i j \gamma] |-1, j\rangle
+
t \sum_{j=1}^{j=n} \exp[i j \gamma] |+1, j\rangle
\Big\}.
\end{equation}
%\end{widetext}
Now, defining $|\Psi_{\mbox{\scriptsize scat.}}\rangle
= \lim_{n \rightarrow + \infty} |\Psi(n)\rangle$, one finds
%\begin{widetext}
\begin{equation}
|\Psi_{\mbox{\scriptsize scat.}}\rangle = \frac{1}{\sqrt{2 \pi}}
\Big\{
\sum_{j=-\infty}^{j=0} \exp[i j \gamma] |+1, j\rangle
+ r
\sum_{j=-\infty}^{j=-1} \exp[-i j \gamma] |-1, j\rangle
+ t \sum_{j=+1}^{j=+\infty}\exp[i j \gamma] |+1, j\rangle
\Big\}.
\label{scatteringonerw}
\end{equation}
\end{widetext}
Note that $U |\Psi_{\mbox{\scriptsize scat.}}\rangle =
|\Psi_{\mbox{\scriptsize scat.}}\rangle$, so it is a stationary state.

An equivalent situation for the generalized Kronig-Penney lattice 
is to assume that all $r$'s but one (the reflection amplitude for 
the point interaction at the origin) are identically null, namely,
$r_j^{(+)} = 0$ and $t_j^{(+)} = 1$ ($j \neq 0$) and 
$r_0^{(+)}(p) = r(p)$, $t_0^{(+)}(p) = t(p)$.
In this case, the scattering solution for a particle incident from 
the left reads
\begin{widetext}
\begin{equation}
|\Phi_{\mbox{\scriptsize scat.}}\rangle = \frac{1}{\sqrt{2 \pi}}
\Big\{
\int_{-\infty}^{0} dx \, \exp[i p x] | x \rangle
+ r(p)
\int_{-\infty}^{0} dx \, \exp[-i p x] | x \rangle
+ t(p) \int_{0}^{+\infty} dx \, \exp[i p x] | x \rangle
\Big\}.
\label{scatteringonefree}
\end{equation}
\end{widetext}
Comparing Eqs. (\ref{scatteringonerw}) and (\ref{scatteringonefree}),
it is evident the correspondence between the two situations.

We can go further, considering that only at two sites the walk can 
``choose'' (from $r_{\pm 1,j}$ and $t_{\pm 1,j}$, $j = 0, \ 1$) a direction 
to proceed, 
whereas at the other sites the direction is always maintained, with 
$\rho_j = 1$ and $\phi_j = 0$ for any $j \neq 0, \, 1$.
Thus, repeating the same calculations for the initial state 
$|\Psi(0)\rangle$ of Eq. (\ref{initial}), we get
\begin{widetext}
\begin{eqnarray}
|\Psi_{\mbox{\scriptsize scat.}} \rangle &=&
\frac{1}{\sqrt{2 \pi}} \Big\{
\sum_{j=-\infty}^{j=0} \exp[i j \gamma] |+1, j\rangle
+ r
\sum_{j=-\infty}^{j=-1} \exp[-i j \gamma] |-1, j\rangle
+ t \sum_{j=+2}^{j=+\infty}\exp[i j \gamma] |+1, j\rangle
\nonumber \\
& & + a |-1, 0\rangle + b \exp[i \gamma] |+1, 1\rangle \Big\},
\end{eqnarray}
%\end{widetext}
where
\begin{equation}
r = r_{+1,0} + t_{-1,0} \, a, \qquad
t = t_{+1,1} \, b, \qquad
a = \frac{t_{+1,0} \, r_{+1,1} \exp[2 i \gamma]}
{1 - r_{+1,1} r_{-1,0} \exp[2 i \gamma]}, \qquad
b = \frac{t_{+1,0}}
{1 - r_{+1,1} r_{-1,0} \exp[2 i \gamma]}.
\label{coef2}
\end{equation}
%\end{widetext}
This expression should be compared with that for the associated
problem of two general point interactions located at
$x=0$ and $x = 1$, whose scattering state (incoming 
from the left) is given by
%\begin{widetext}
\begin{eqnarray}
|\Phi_{\mbox{\scriptsize scat.}}\rangle &=&
\frac{1}{\sqrt{2 \pi}}
\Big\{
\int_{-\infty}^{0} dx \, \exp[i p x] | x \rangle
+ r(p)
\int_{-\infty}^{0} dx \, \exp[-i p x] | x \rangle
+ t(p) \int_{1}^{+\infty} dx \, \exp[i p x] | x \rangle
\nonumber \\
& & + a(p) \int_{0}^{1} dx \exp[-i p x] | x \rangle +
b(p) \int_{0}^{1} dx \exp[i p x] | x \rangle \Big\},
\end{eqnarray}
\end{widetext}
for the coefficients $r(p)$, $t(p)$, $a(p)$ and $b(p)$ 
obtained from Eq. (\ref{coef2}) through the substitutions
$r_{\pm 1,j} \rightarrow r_j^{(\pm)}(p)$,
$t_{\pm 1,j} \rightarrow t_j^{(\pm)}(p)$ and
$\gamma \rightarrow p$.
Once more we find a direct association between the two cases.

By repeating this procedure of ``turning on'' more and more sites 
in the quantum walk and zero-range potentials in the Kronig-Penney 
lattice, one realizes that their relation is indeed direct.
The one-to-one mapping is a simple identification of quantities 
in the two cases. 
The direction coefficients $r_{\sigma, j}$ and $t_{\sigma, j}$  
at each site in the quantum walk corresponds to the scattering 
amplitudes $r_{j}^{(\pm)}$ and $t_{j}^{(\pm)}$ of a point interaction in 
the Kronig-Penney model.
The quantum number $j$ is associated to the appropriate position 
eigenvalues $x/L$, likewise for $\sigma$ with respect to the 
signal of $p$.
Lastly, the SQW single step evolution $U$ (up to the phase $\exp[i \gamma]$)
is akin to ${\mathcal U}{(t = \tau)}$ for the continuous scattering system.

We finally note that we have discussed the mapping assuming a 
scattering scenario, with the QW initial state given by Eq. 
(\ref{initial}).
However, we also could start with an initial state 
localized in some region of the quantum walk lattice and an initial 
wave packet localized in an equivalent region of the generalized 
Kronig-Penney lattice.
Then, by applying the respective time evolution operators, again
we would find a direct association between their dynamics: 
the multiple scattering of the wave packet in the Kronig-Penney lattice 
resembling the proliferation of paths (e.g., see the example in Sec. 
II) in the quantum walk.
So, the correspondence between the two systems is complete and not 
restricted to the type of initial state assumed.
This fact becomes more evident from the Green function approach next.

\subsection{Quantum walks and finite lattices}

\begin{figure}
  \centering
\centerline{\psfig{figure=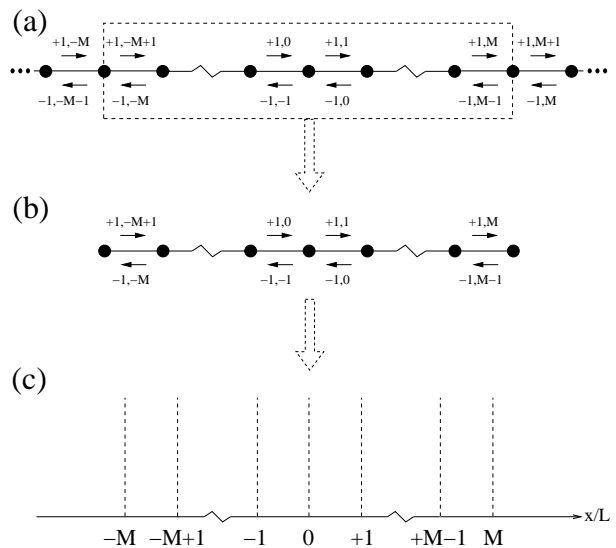,width=8cm}}
  \caption{
    (a) If under a particular instance, the QW relevant dynamics 
    is restricted to the 
    states $|j| \leq M$; (b) then, effectively the system 
    can be described by a finite Hilbert space ``lattice''; 
    (c) whose mapping leads to a finite set of general point 
    interactions on the line.}
  \label{fig3}
\end{figure}

To calculate the exact Green function in the case of an infinite 
generalized Kronig-Penney lattice is a difficult task
\cite{schmidt1}. 
However, a key aspect in solving QW through the 
proposed mapping is that in almost all situations of interest,
the original system can be mapped to a finite lattice -- a 
limited number of point scatters along the line -- and not to an 
infinite comb of zero-range potentials.

For example, let us assume that the quantum walk initial state
$|\Psi(0)\rangle$ is, say, either $|-1,-1\rangle$ or $|+1,+1\rangle$, 
thus localized about and leaving from the origin.
Now, suppose we shall discuss any quantity for times no longer than 
$n=N$, or for contexts where the dynamics never takes the system
beyond the sites $j = \pm J$, $J > 0$.
Examples are:
(a) to determine the probability to be at the state $j$ (i.e., 
to calculate $|\langle j, \sigma | \Psi(n)\rangle|^2$) for $n$ 
up to $n=N$; 
and
(b) to obtain the probability for the walk to reach for the very
first time a ``distance'' $j=J$ from the origin ($j = 0$) at times
$n = 1, 2, \ldots$, known as the first passage time problem in 
classical random walk theory \cite{Book-Rudnick}.

For (b), any evolution leading to 
$\langle j, \sigma \, | U^n \, | \Psi(0)\rangle \neq 0$
($|j| > J$, arbitrary $n$) has no interest for the problem solution
\cite{kempe}.
In (a), after $N$ steps the initial state has spread at most a 
distance $|j| = N$ from the origin.
Hence, as illustrated in Fig. 3, in both situations the relevant 
dynamics for the QW is related just to a segment of the infinite 
generalized Kronig-Penney lattice, encompassing 
$2 M + 1$ (for $M$ equal to $J$ or $N$) point interactions.
So, in such instances effectively one needs to deal only with 
finite lattices.

\subsection{The finite lattice Green function and its 
relation to the original quantum walk problem}

Once the quantum walk dynamics one shall study is mapped to an appropriate 
(finite) generalized Kronig-Penney lattice, the next step is to
calculate the Green function for the latter.
Based on certain techniques \cite{luz,fabiano-tese}, the way 
to do so has been developed in \cite{schmidt1}.
Here we just summarize the main steps (for details see \cite{schmidt1}).

Suppose a particle of energy $E = k^2/2$, for which $G(x_f, x_i; k)$
denotes its energy-dependent Green function.
The initial and final positions, respectively $x_i$ and $x_f$, are
arbitrary points along the 1D lattice (e.g., Fig. 4).
Then, the exact $G$ (up to a factor $(i k)^{-1}$, unnecessary for
our purposes) 
reads \cite{luz,schmidt1}
\begin{equation}
\label{green}
G(x_f, x_i; k) = \sum_{\mbox{\scriptsize
s.p.}} W_{\mbox{\scriptsize s.p.}} \exp[i S{\mbox{\scriptsize
s.p.}}(x_f,x_i;k)].
\end{equation}
The sum is performed over all possible infinite ``scattering paths'' 
(s.p.) starting and ending at the points $x_i$ and $x_f$.
For each s.p., the classical action is written as 
$S_{\mbox{\scriptsize s.p.}} = k \, L_{\mbox{\scriptsize s.p.}}$, 
with $L_{\mbox{\scriptsize s.p.}}$ the s.p. total length.
The pre-factor amplitude (or weight) $W_{\mbox{\scriptsize s.p.}}$ 
is given by the product of the quantum coefficients gained each 
time the particle is scattered off by a given contact potential 
along the way.

\begin{figure}
  \centering
\centerline{\psfig{figure=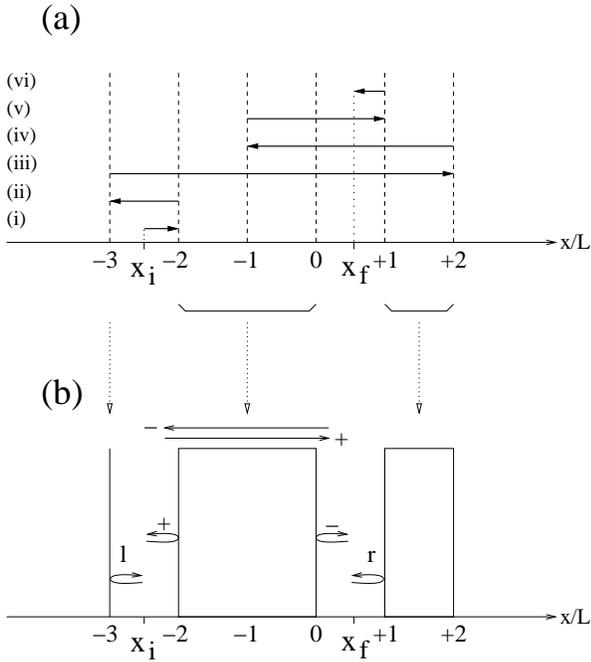,width=8cm}}
  \caption{(a) For a finite Kronig-Penney lattice of six general 
point interactions, and specific end points $x_i$ and $x_f$,
a representative ``scattering path'' composed by stretches, 
(i)--(vi), of straight trajectories.
It has a total length of 
$L_{\mbox{\scriptsize s.p.}} = 11 + (-2 - x_i) + (1 - x_f)$, ($L = 1$).
(b) The ${\mathcal R}$'s and ${\mathcal T}$'s in Eq. 
(\ref{green-example}) are the resulting composite reflection and 
transmission amplitudes for sets of point interaction potentials, 
as illustrated.}
  \label{fig4}
\end{figure}

To illustrate a typical term in Eq. (\ref{green}), we consider a 
lattice with six point interactions equally spaced by $L = 1$.
Taking as the end points $-3 < x_i < -2$ and $0 < x_f < 1$, a
representative scattering path is depicted in Fig. 4. 
For such s.p., the particle starts at $x_i$, goes to the right, 
reflects from $x=-2$, moves to the left, reflects from $x=-3$, and 
then goes to the right, tunneling all the potentials until 
reflecting from $x=2$.
In this part of the trajectory -- stretches (i), (ii), and (iii) 
in Fig. 4 (a) -- the partial weight is 
$W_{(i)+(ii)+(iii)} = r_{-2}^{(+)} \, r_{-3}^{(-)} \, t_{-2}^{(+)}
\, t_{-1}^{(+)} \, t_{0}^{(+)}\, t_{+1}^{(+)}\, r_{+2}^{(+)}$.
From $x = 2$, the particle travels to the left, is transmitted
through the potentials at $x=1$ and $x=0$, and then is reflected 
by the point interaction at $x=-1$.
Next, it travels to $x=1$ (tunneling the potential at the origin), 
suffers another reflection, and finally gets to the end point $x_f$.
In this part of the trajectory -- (iv), (v), and (vi) in Fig. 4 (a)
-- the amplitude is
$W_{(iv)+(v)+(vi)} = t_{+1}^{(-)} \, t_{0}^{(-)} \, r_{-1}^{(-)} \,
t_{0}^{(+)} \, r_{+1}^{(+)}$.
Hence, the total pre-factor weight for this particular s.p. is
$W_{\mbox{\scriptsize s.p.}} = W_{(i)+(ii)+(iii)} \times W_{(iv)+(v)+(vi)}$.
The scattering path length is simply
$L_{\mbox{\scriptsize s.p.}} = 11 + (-2-x_i) + (1-x_f)$, as readily 
seen from Fig. 4 (a).

To obtain $G$ in a closed form, one should classify and to sum up 
(c.f., Eq. (\ref{green})) all the infinitely many possible trajectories 
of the kind exemplified above.
Fortunately, it always can be done by regrouping the infinite sets of 
trajectories into a finite number of classes 
\cite{dyson,schmidt1}.
Furthermore, as proved in \cite{luz}, these classes form geometric 
series, allowing their exact summation.
For instance, from such procedure the correct Green function for
the system in Fig. 4 can be calculated, leading to \cite{luz,schmidt1}
\begin{widetext}
\begin{equation}
G(x_f, x_i ; k) = 
\frac{\mathcal{T}_{+}}{[1 -  \mathcal{R}_{\mbox{\scriptsize l}} \mathcal{R}_{+}]
                   [1 - \mathcal{R}_{-} \mathcal{R}_{\mbox{\scriptsize r}}] 
      - \mathcal{T}_{+} \mathcal{T}_{-} \mathcal{R}_{\mbox{\scriptsize l}}
                      \mathcal{R}_{\mbox{\scriptsize r}}}
\Big(\exp[- i k x_i] + \mathcal{R}_{\mbox{\scriptsize l}} \exp[i k x_i]\Big)
\Big(\exp[i k x_f] + \mathcal{R}_{\mbox{\scriptsize r}} \exp[- i k x_f]\Big).
\label{green-example}
\end{equation}
In the above expression, the ${\mathcal R}$'s and ${\mathcal T}$'s
are effective reflection and transmission amplitudes, resulting from
groups of zero range potentials as depicted in Fig. 4 (b). 
They are explicit given by
\begin{eqnarray}
\mathcal{R}_{\mbox{\scriptsize l}} &=& r_{-3}^{(-)} \exp[6 i k], 
\nonumber \\
\mathcal{R}_{\mbox{\scriptsize r}} &=& r_{+1}^{(+)} \exp[2 i k]
+ \frac{t_{+1}^{(-)} t_{+1}^{(+)} \, r_{+2}^{(+)} \exp[4 i k]}
       {1 - r_{+1}^{(-)} r_{+2}^{(+)} \exp[2 i k]},
\nonumber \\ 
\mathcal{R}_+ &=& r_{-2}^{(+)} \exp[- 4 i k]
+ \frac{
\left(
      r_{-1}^{(+)} - 
       \left(r_{-1}^{(-)} r_{-1}^{(+)} - t_{-1}^{(-)} t_{-1}^{(+)} \right) 
       r_{0}^{(+)}  \exp[2 i k] \right) 
       t_{-2}^{(-)} t_{-2}^{(+)} \exp[-2 i k] }
    {1-
      \left(r_{-2}^{(-)} r_{-1}^{(+)}+r_{-1}^{(-)} r_{0}^{(+)}\right)
      \exp[2 i k] +
      \left(r_{-1}^{(-)} r_{-1}^{(+)}-t_{-1}^{(-)} t_{-1}^{(+)}\right)
      r_{-2}^{(-)} r_{0}^{(+)} \exp[4 i k]},
\nonumber \\
  \mathcal{T}_+ &=&
  \frac{t_{-2}^{(+)} t_{-1}^{(+)} t_{0}^{(+)}}
{1-
      \left(r_{-2}^{(-)} r_{-1}^{(+)}+r_{-1}^{(-)} r_{0}^{(+)}\right)
      \exp[2 i k] +
      \left(r_{-1}^{(-)} r_{-1}^{(+)}-t_{-1}^{(-)} t_{-1}^{(+)}\right)
      r_{-2}^{(-)} r_{0}^{(+)} \exp[4 i k]},
\nonumber \\
\mathcal{R}_- &=& \mathcal{R}_+ \exp[4 i k], \ \mbox{where in} \
\mathcal{R}_+ \ \mbox{we exchange} \ \
(+) \longleftrightarrow (-) \ \mbox{and} \ j=-2 \longleftrightarrow j=0, 
\nonumber \\
  \mathcal{T}_- &=& \mathcal{T}_+, \ \mbox{where in} \
\mathcal{T}_+  \ \mbox{we exchange} \ \
(+) \longleftrightarrow (-) \ \mbox{and} \ j=-2 \longleftrightarrow j=0. 
\label{green-example-coef}
\end{eqnarray}
\end{widetext}

Here, two points should be emphasized: 
(a) the energy domain $G$ is given by a sum over all the possible 
multiple scattering processes suffered by the particle; 
(b) each s.p. in the series Eq. (\ref{green}) represents a
trajectory in which the particle spends a time of roughly
$t \sim n \tau$, for $n$ the number of scattering along the path 
(e.g., $n=12$ in the example of Fig. 4).

In the study of QW, common questions are related to the system state, 
say, after evolving $n$ steps.
But from the (a)-(b) above, such information is fully contained in 
the series representation of $G$.
% unveiling the interference character of QW.
Therefore, since the correspondence between QW and generalized 
Kronig-Penney lattices is straightforward, we can readily associate 
each term in Eq. (\ref{green}) to a possible evolution of a quantum walk
(e.g., that in Eq. (\ref{example-3steps})).
Moreover, such terms can be viewed as a Fourier decomposition of $G$. 
Given that the Green function Fourier transform is the time domain 
propagator, an individual term in Eq. (\ref{green}), when 
properly mapped, represents then a possible path for $t = n$ time steps 
in the quantum walk.

Finally, depending on specific QW quantities we shall calculate, in 
practice a simple inspection and selection of paths in the expansion 
for $G$ will suffice.
However, for larger $n$'s and more complicated topologies
(Sec. IV), it may be cumbersome to deal with individual terms 
in Eq. (\ref{green}).
Fortunately, one can make the Green function a systematic protocol
for QW by introducing the {\em path and step} operators.
As we discuss in Sec. V, they are useful tools to extract any 
information about the system directly from an already summed 
closed expression for $G$.

\section{Extension to arbitrary topologies}

\begin{figure}
  \centering
\centerline{\psfig{figure=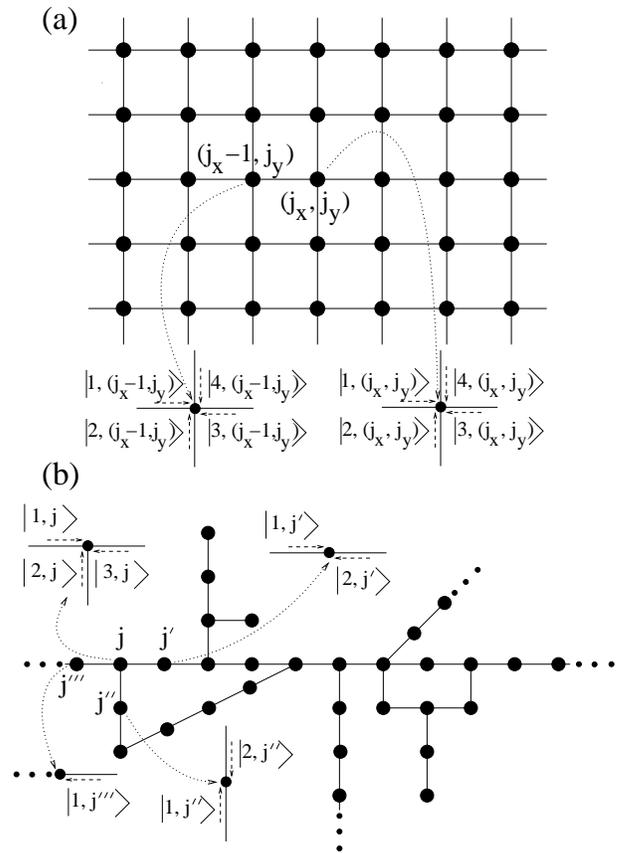,width=8cm}}
  \caption{Examples of graph structures, which generalize 1D QW.
    For SQW, the states (as illustrated) are 
    defined on the bonds joining the different sites $j$.
    (a) All the sites have a same number of first neighbors in a
        regular topology.
    (b) For an irregular structure, such number depends on $j$.}
  \label{fig5}
\end{figure}

QW can be defined in arbitrary topologies \cite{kempe}, 
i.e., for general graph structures \cite{andrade}.
The scattering formulation is then obtained through a direct extension 
of the construction in Sec. II \cite{hillery-pra,feldman-jpa,andrade}.

Suppose an undirected simple network \cite{kendon-topology} of sites 
connected by bonds (examples in Fig. 5).
Its topology represents the particular Hilbert space arrangement in 
which the quantum walk dynamics takes place.
Like the 1D lattice, each bond joining two neighbor sites 
-- say $j$ and $j'$ -- is associated only two basis states, one 
incoming to $j$ and other to $j'$.
For instance, for the bond joining $(j_x-1, j_y)$ to $(j_x, j_y)$ 
in Fig. 5 (a), we have $|1, (j_x,j_y) \rangle$ and 
$|3, (j_x-1, j_y) \rangle$, whereas for the bond connecting $j$ to
$j'$ ($j$ to $j''$) in Fig. 5 (b), we have $|3, j \rangle$ and 
$|1, j' \rangle$ ($|2, j \rangle$ and $|2, j'' \rangle$).
But contrary to the 1D case, the possible ``directions'' (bonds)
to get to a site $j$ from its first neighbors may depend on the 
specific $j$.
Thus, the quantum number $\sigma_j$ assumes the values 
$1, 2, \ldots, K_j$, with $K_j$ the coordination number (valence) 
of site $j$.
In more regular structures $K_j = K$ regardless of $j$ (e.g., 
$K=4$ in Fig. 5 (a)).

The construction of the time evolution operators is discussed, e.g., 
in Refs. \cite{hillery-pra,feldman-jpa,andrade}.
Here we just outline the main ideas following Ref. \cite{andrade}.
First, one needs to characterize the lattice topology, namely,
to specify for any $j$ the sets:
(a) ${\cal{S}}_j = \{j^{(1)},j^{(2)},\ldots,j^{(K_j)}\}$ of all the 
$K_j$ sites which are first neighbors of $j$ (e.g., in Fig. 
5 (b), ${\cal{S}}_j = \{j', j'',j'''\}$); 
(b) ${{\cal N}}_j = \{\sigma_{j^{(1)}}, \sigma_{j^{(2)}}, \ldots, 
\sigma_{j^{(K_j)}} \}$ for $\sigma_{j^{(n)}}$ the direction quantum number 
for the state incoming to site $j^{(n)}$ through the bond joining 
$j^{(n)}$ and $j$ (in Fig. 5 (b), ${\cal{N}}_j = 
\{\sigma_{j'} = 1, \sigma_{j''} = 2, \sigma_{j'''} = 1\})$; and
(c) %${\cal{B}}_j = \{ \sigma_{j}^{(1)}, \sigma_{j}^{(2)}, \ldots, 
%\sigma_{j}^{(K_j)} \}$, 
${\cal{B}}_j = \{ \tilde{\sigma}_{j^{(1)}}, 
\tilde{\sigma}_{j^{(2)}}, \ldots, \tilde{\sigma}_{j^{(K_j)}} \}$
%with $\sigma_{j}^{(n)}$ the direction quantum
with $\tilde{\sigma}_{j^{(n)}}$ the direction quantum
number for the state 
%$|\sigma_{j}^{(n)}, j \rangle$ incoming to
$|\tilde{\sigma}_{j^{(n)}}, j \rangle$ incoming to
$j$ along the bond connecting $j$ and $j^{(n)}$ (in Fig. 5 (b),
%${\cal{B}}_j = \{ 
%\sigma_{j}^{(1)} = 3: \, \mbox{bond} \ j\mbox{--}j', \ 
%\sigma_{j}^{(2)} = 2: \, \mbox{bond} \ j\mbox{--}j'', \ 
%\sigma_{j}^{(3)} = 1: \, \mbox{bond} \ j\mbox{--}j''' \}$).
${\cal{B}}_j = \{ 
\tilde{\sigma}_{j^{(1)}} = 3: \, \mbox{bond} \ j\mbox{--}j', \ 
\tilde{\sigma}_{j^{(2)}} = 2: \, \mbox{bond} \ j\mbox{--}j'', \ 
\tilde{\sigma}_{j^{(3)}} = 1: \, \mbox{bond} \ j\mbox{--}j''' \}$).

Second, one defines the one step time evolution $U$ in terms of its
action over each basis state $|\sigma_j, j \rangle$, or
(with $\sigma_j = \tilde{\sigma}_{j^{(i)}} \in {\cal{B}}_j$ and $\sigma_{j^{(i)}}$
the corresponding element in ${{\cal N}}_j$)
\begin{eqnarray}
U(\gamma) |\sigma_j, j \rangle &=& \exp[i \gamma] \Big(
r_{\sigma_j \sigma_{j}, j} |\sigma_{j^{(i)}}, j^{(i)} \rangle 
\nonumber \\
& & + 
\sum_{n=1; n \neq i}^{n = K_j} t_{\sigma_j \tilde{\sigma}_{j^{(n)}}, j} 
|\sigma_{j^{(n)}}, j^{(n)} \rangle \Big).
\label{u-topology}
\end{eqnarray}
Finally, the $r$'s and $t$'s are chosen such that for any $j$ 
the $K_j \times K_j$ matrix $\Gamma_j$ (of elements 
$[\Gamma_j]_{\sigma \, \sigma} = r_{\sigma \sigma, j}$ and
$[\Gamma_j]_{\sigma' \, \sigma} = t_{\sigma' \sigma, j}$, for both 
$\sigma \neq \sigma'$ in $\{1, 2, \ldots, K_j\}$) is unitary.
This makes $U$ also unitary \cite{andrade}, establishing SQW in 
arbitrary topologies.

The usual, i.e., continuous in time and space, quantum mechanical 
dynamics on network structures (known as quantum graph systems 
\cite{kottos1}) is likewise a generalization of the evolution in a 
1D lattice with zero range potentials \cite{schmidt1,kottos2}.
It is obtained by properly matching the solutions of a 1D free 
\cite{comment2} Schr\"odinger equation in the different arms (bonds) 
at the vertices (sites).
For this end, one assumes for each $j$ a matrix $S_j(k)$ (see below), 
which describes how an incoming plane wave of wave number $k$ is 
scattered off at the vertex $j$. 
So, any $j$ can be viewed as a general point interaction, but
connecting $K_j$-directions instead of the common two (left and right) 
on the line.
Furthermore, if for all $j$, $S_j {S_j}^\dag = {S_j}^\dag S_j = {\bf 1}$,
the resulting dynamics is unitary, conserving flux probability.

Quantum graphs can be solved through a Green function approach 
\cite{barra}.
In fact, it has been shown \cite{schmidt2} that the exact $G$ is
also given by Eq. (\ref{green}), where now the scattering paths are
all the possible trajectories along the network, starting and ending 
at the points $x_i$ and $x_f$ (located in arbitrary arms of the graph).
The $W$'s are the quantum amplitudes gained along the s.p.'s due 
to the scattering at the different sites. 
Finally, the mentioned procedure of classifying and summing up 
different classes of s.p.'s still holds in this case \cite{schmidt2}.
So, we always can write the exact $G$ as a closed analytical 
expression.

Summarizing, QW in general networks are direct extensions of QW 
in 1D exactly in the same way than quantum graphs are natural 
extensions of Kronig-Peney lattices.
Therefore, it is easy to realize that our previous mapping between 
the two types of systems in 1D remains valid in arbitrary topologies 
too.

Lastly, to define the reflection and transmission scattering 
amplitudes in a quantum graph -- and to associate them to QW
coefficients -- we consider the same labeling used to
characterize the lattices topologies.
Thus, for 
%${\cal{B}}_j = 
%\{ \sigma_{j}^{(1)}, \sigma_{j}^{(2)}, \ldots, \sigma_{j}^{(K_j)} \}$,
${\cal{B}}_j =
\{ \tilde{\sigma}_{j^{(1)}}, \tilde{\sigma}_{j^{(2)}}, \ldots, 
\tilde{\sigma}_{j^{(K_j)}} \}$,
the matrix elements of $S_j$ are
\begin{equation}
%[S_j]_{i \, i} =  r_j^{(\sigma_{j}^{(i)} \sigma_{j}^{(i)})}, \qquad
%[S_j]_{i \, l} =  t_j^{(\sigma_{j}^{(i)} \sigma_{j}^{(l)})} \ (i \neq l).
[S_j]_{i \, i} =  r_j^{(\tilde{\sigma}_{j^{(i)}} \tilde{\sigma}_{j^{(i)}})}, \qquad
[S_j]_{i \, l} =  t_j^{(\tilde{\sigma}_{j^{(i)}} \tilde{\sigma}_{j^{(l)}})} \ (i \neq l).
\label{s-matrix}
\end{equation}
In Eq. (\ref{s-matrix}), 
%$r_j^{(\sigma_{j}^{(i)} \sigma_{j}^{(i)})}$ 
%($t_j^{(\sigma_{j}^{(i)} \sigma_{j}^{(l)})}$) 
$r_j^{(\tilde{\sigma}_{j^{(i)}} \tilde{\sigma}_{j^{(i)}})}$ 
($t_j^{(\tilde{\sigma}_{j^{(i)}} \tilde{\sigma}_{j^{(l)}})}$) 
is the reflection (transmission)
coefficient for the particle incoming to site $j$ from bond 
%$\sigma_{j}^{(i)}$ 
$\tilde{\sigma}_{j^{(i)}}$ to be reflected (transmitted) to bond 
%$\sigma_{j}^{(i)}$ ($\sigma_{j}^{(l)}$). 
$\tilde{\sigma}_{j^{(i)}}$ ($\tilde{\sigma}_{j^{(l)}}$).
The unitarity of the $S_j$'s plus the symmetries of 
the Schr\"odinger equation for real potentials \cite{chadan} 
(i.e., ${S_{j}}^{\dag}(k) = S_{j}(-k)$), yield (where 
$i, l, n = 1, 2, \ldots, K^{(j)}$)
% and the script $j$ is dropped for simplicity of notation) 
%\begin{eqnarray}
%& & 
%r^{(\sigma^{(i)} \sigma^{(i)})}(k) = {r^{(\sigma^{(i)} \sigma^{(i)})}}^*(-k), 
%\nonumber \\
%& & 
%t^{(\sigma^{(i)} \sigma^{(l)})}(k) = {t^{(\sigma^{(l)} \sigma^{(i)})}}^*(-k),
%\nonumber \\
%& & 
%\sum_{l \neq i} t^{({\sigma^{(i)} \sigma^{(l)}})}  
%{t^{({\sigma^{(i)} \sigma^{(l)}})}}^* + 
%r^{(\sigma^{(i)} \sigma^{(i)})}  {r^{(\sigma^{(i)} \sigma^{(i)})}}^* = 1,
%\nonumber \\
%& & 
%\sum_{n \neq i,l} t^{({\sigma^{(i)} \sigma^{(n)}})}  
%{t^{({\sigma^{(l)} \sigma^{(n)}})}}^* +
%r^{(\sigma^{(i)} \sigma^{(i)})} {t^{({\sigma^{(l)} \sigma^{(i)}})}}^* 
%\nonumber \\
%& & + 
%{r^{(\sigma^{(l)} \sigma^{(l)})}}^* t^{({\sigma^{(i)} \sigma^{(l)}})} = 0.
%\label{qamplitudes}
%\end{eqnarray}
\begin{eqnarray}
& & 
r_j^{(\tilde{\sigma}_{j^{(i)}} \tilde{\sigma}_{j^{(i)}})}(k) = 
{r_j^{(\tilde{\sigma}_{j^{(i)}} \tilde{\sigma}_{j^{(i)}})}}^*(-k), 
\nonumber \\
& & 
t_j^{(\tilde{\sigma}_{j^{(i)}} \tilde{\sigma}_{j^{(l)}})}(k) = 
{t_j^{(\tilde{\sigma}_{j^{(l)}} \tilde{\sigma}_{j^{(i)}})}}^*(-k),
\nonumber \\
& & 
\sum_{l \neq i} t_j^{({\tilde{\sigma}_{j^{(i)}} \tilde{\sigma}_{j^{(l)}}})}  
{t_j^{({\tilde{\sigma}_{j^{(i)}} \tilde{\sigma}_{j^{(l)}}})}}^* + 
r_j^{(\tilde{\sigma}_{j^{(i)}} \tilde{\sigma}_{j^{(i)}})}  
{r_j^{(\tilde{\sigma}_{j^{(i)}} \tilde{\sigma}_{j^{(i)}})}}^* = 1,
\nonumber \\
& & 
\sum_{n \neq i,l} t_j^{({\tilde{\sigma}_{j^{(i)}} \tilde{\sigma}_{j^{(n)}}})}  
{t_j^{({\tilde{\sigma}_{j^{(l)}} \tilde{\sigma}_{j^{(n)}})}}}^* +
r_j^{(\tilde{\sigma}_{j^{(i)}} \tilde{\sigma}_{j^{(i)}})} 
{t_j^{({\tilde{\sigma}_{j^{(l)}} \tilde{\sigma}_{j^{(i)}}})}}^* 
\nonumber \\
& & + 
{r_j^{(\tilde{\sigma}_{j^{(l)}} \tilde{\sigma}_{j^{(l)}})}}^* 
t_j^{({\tilde{\sigma}_{j^{(i)}} \tilde{\sigma}_{j^{(l)}}})} = 0.
\label{qamplitudes}
\end{eqnarray}
The above are natural generalizations \cite{chadan,schmidt2} of the usual 
relations for the scattering coefficients (c.f., Eqs. 
(\ref{condition-coef}) and (\ref{rr-tt})) of a point scatterer on 
the line.
Note also that if we impose time-reversal invariance, 
%$t^{({\sigma^{(i)} \sigma^{(l)}})} = t^{({\sigma^{(l)} \sigma^{(i)}})}$.
$t_j^{({\tilde{\sigma}_{j^{(i)}} \tilde{\sigma}_{j^{(l)}}})} = 
t_j^{({\tilde{\sigma}_{j^{(l)}} \tilde{\sigma}_{j^{(i)}}})}$.

Hence, the direction coefficients in a quantum walk, the $\Gamma_j$'s, 
are in one-to-one correspondence with the scattering matrices $S_j$'s
in a quantum graph system.

\section{Extracting information from $G$: the step and path operators}

From the previous results, it turns out that
the exact Green function, Eq. (\ref{green}), is actually 
the {\em generating function} of all the possible walks 
leaving from and arriving at the bonds corresponding to $x_i$ 
and $x_f$, respectively. 
So, any individual or group of QW paths are obtained through 
proper manipulations of $G$.

In this way, more simple tasks like to determine all the trajectories 
for $|\Psi(0)\rangle = |\sigma, j\rangle$ evolving, say, 
only two times steps ($n=2$), can be done by identifying particular 
terms directly in the $G$ series representation, Eq. (\ref{green}). 
However, the huge proliferation of paths in instances such as to 
find certain trajectories connecting two bonds very far apart, 
or resulting from high values of $n$, makes the full series 
expansion difficult to deal with.
In such cases, a better approach is first to sum up the series 
\cite{bach} (using the already mentioned procedures in the literature
\cite{luz,schmidt1,schmidt2,fabiano-tese} to get 
expressions like Eq. (\ref{green-example})) and then to extract 
the sought information from $G$ with the help of the two operators 
described below.

The first is $\hat{S}_n$, yielding all the paths of exactly 
$n$ time steps.
To define $\hat{S}_n$, note that any walk state gains a factor 
$\exp[i \gamma]$ at each time step (see Eq. (\ref{u-topology})).
From the mapping, such factor is equivalent to $z = \exp[i k L]$ in 
the (continuous) quantum graph problem.
So, let us set $G_z$ as $G$ in Eq. (\ref{green}), but with the
substitution $\exp[i k L] \rightarrow z$, and for which the 
scattering amplitudes are identified with the appropriate quantum walk 
coefficients $\Gamma_j$'s (Section IV).
Thus, one finds that if the step operator $\hat{S}_n$ acting on $G_z$ 
has the form 
\begin{equation}
\hat{S}_n \equiv \frac{1}{n!}
\frac{\partial^n}{\partial z^n}\bigg|_{z=0},
\end{equation}
then $\big|\hat{S}_n G_z\big|^2$ gives the total probability for 
the quantum walk to leave the bond $x_i$ and to get to the bond 
$x_f$ in exactly $n$ steps.
We should mention that such construction has already been proposed 
in \cite{feldman-pla,feldman-jpa} to treat scattering problems.
Considering the Green function approach, we see that 
$\hat{S}_n$ can be applied in more general contexts.

The second is $\hat{P}_{\mathcal P}$, which extracts from $G$ all the
paths with specific trajectory stretches ${\mathcal P}$.
Any quantum walk s.p. can be described by the sequence of 
coefficients $r$ and $t$ it gains along the way (cf., Eq. 
(\ref{example-3steps})).
For instance, consider Fig. 5 (b) and assume $n=6$ applications of 
$U$ to the system initially at $|1,j  \rangle$.
One possible sequence of successively visited states during the 
evolution, thus representing a possible path, is:
\begin{equation}
|1,j  \rangle {\small \rightarrow} |1,j'  \rangle {\small \rightarrow} 
|3,j  \rangle {\small \rightarrow} |1,j'  \rangle {\small \rightarrow} 
|3,j  \rangle {\small \rightarrow} |2,j'' \rangle {\small \rightarrow} 
|2,j  \rangle.
\nonumber
\end{equation}
Here $W = t_{1 \, 1, j} \, r_{1 \, 1, j'} \, r_{3 \, 3, j} \, 
r_{1 \, 1, j'} \, t_{3 \, 2, j} \, r_{2 \, 2, j''}$ is its
probability amplitude, which can be rewrite as 
$W = (r_{1 \, 1, j'})^2 \, (r_{3 \, 3, j})^1 \, (r_{2 \, 2, j''})^1 \,
(t_{1 \, 1, j})^1 \, (t_{3 \, 2, j})^1$.
Thus, any trajectory (or part of it) can be represented by 
$\mathcal{P} = \{(\alpha_1,n_{\alpha_1}),(\alpha_2,n_{\alpha_2}), \ldots;
(\beta_1,n_{\beta_1}),(\beta_2,n_{\beta_2}), \ldots\}$, i.e., by the
set of indexes and exponents associated to the quantum coefficients 
$r$ and $t$ of the path stretch (with $\alpha$ and $\beta$
standing for  $\sigma' \, \sigma'', j$).
In our example,
$\alpha_1 = 1 \, 1, j'$; $\alpha_2 = 3 \, 3, j$; $\alpha_3 = 2 \, 2, j$;
$\beta_1 = 1 \, 1, j'$; $\beta_2 = 3 \, 2, j$; 
$n_{\alpha_1} = 2$; $n_{\alpha_2} = n_{\alpha_3} = n_{\beta_1} = n_{\beta_2} = 1$.

Now, by properly choosing $x_i$ and $x_f$ (which obviously depends on 
${\mathcal P}$, see Sec. VI) we get that 
$G_{\mathcal P} = \hat{P}_{\mathcal P} G$ is a sum -- in the form of 
Eq. (\ref{green}) -- but containing only paths whose parts of
their trajectories are given by $\mathcal{P}$.
The explicit expression for $\hat{P}_{\mathcal P}$ is
\begin{equation}
\hat{P}_{\mathcal P} \equiv \prod_{\alpha \in {\mathcal P}}
\frac{(r_{\alpha})^{n_{\alpha}}}{n_{\alpha}!} \times
\frac{\partial^{n_{\alpha}}}{\partial r_{\alpha}^{n_{\alpha}}} 
\bigg|_{r_{\alpha}=0}
           \prod_{\beta \in {\mathcal P}}
\frac{(t_{\beta})^{n_{\beta}}}{n_{\beta}!} \times
\frac{\partial^{n_{\beta}}}{\partial t_{\beta}^{n_{\beta}}}
\bigg|_{t_{\beta}=0},
\label{path-operator}
\end{equation}
which must act on the Green function as the following:
first one performs all the indicated derivatives; 
second, one sets the coefficients $r_{\alpha}$ and $t_{\beta}$ 
equal to zero; 
finally one multiplies the resulting expression by the coefficients 
$(r_{\alpha})^{n_{\alpha}}$'s and $(t_{\beta})^{n_{\beta}}$'s.

If we shall select just a path which is itself entirely represented
by $\mathcal{P}$, then in the above definition we simply change 
$|_{r_{\alpha}=0}$ and $|_{t_{\beta}=0}$ by $|_{r=0}$ and $|_{t=0}$, with $r$ and 
$t$ {\em all} the quantum amplitudes in $G$.

Lastly, we note that for an initial state being the superposition 
of $N$ basis states, 
$|\Psi(0)\rangle = \sum_{\sigma, j} c_{\sigma, j} |\sigma, j\rangle$,
we must consider $N$ Green functions, each with a $x_i$ corresponding
to the bond of $|\sigma, j\rangle$.
So, in any calculation, the contribution of each of these $G$'s 
should be weighted by the associated factor $c_{\sigma, j}$.

\section{An example: a diamond-shaped graph}

Finally, to illustrate some features of our Green function approach, 
we discuss a quantum walk for the topology depicted in Fig. 6.
We assume complete general coefficients (observing Eq. 
(\ref{qamplitudes})), in the diamond region -- sites 
$A$, $B$, $C$ and $D$ -- and free evolution, $r_j = 0$ and $t_j = 1$, 
in the leads region -- sites $j \leq -1$ and $j \geq 0$.
We mention that this system, in the case of 
$r_{A} = r_{D} = -1/3$, $t_{A} = t_{D} = 2/3$, 
$r_B = r_C = 0$ and $t_B= t_C = 1$, has been studied in Ref.
\cite{feldman-pla}.
Such particular values represent the so called Grover coins
(see, e.g., \cite{tregenna}).

\begin{figure}
  \centering
\centerline{\psfig{figure=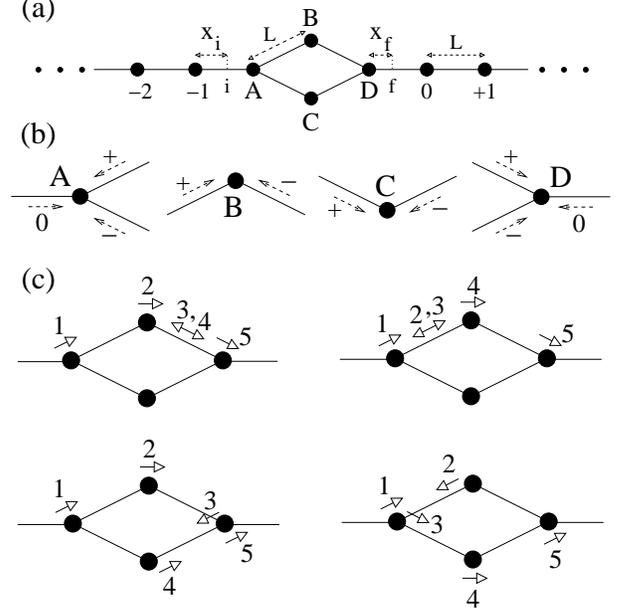,width=8cm}}
  \caption{(a) A graph composed by a diamond-shaped region 
    (sites $A$, $B$, $C$, $D$) attached to semi-infinite 
    leads (sites $j\leq -1$ and $j \geq 0$, for which
    $r_j = 0$ and $t_j = 1$).
    (b) The $\sigma$ labelling for $A$, $B$, $C$ and $D$.
    (c) For $n=5$ time steps, schematics of all possible s.p.'s first
    entering into the diamond region through the superior bond (those
    entering through the inferior bond are simple specular images of (c)).}
  \label{fig6}
\end{figure}

Consider this quantum graph for the initial and the end positions 
$x_i$ and $x_f$, respectively, in the bonds $i$ and $f$, Fig. 6 (a).
The exact Green function reads 
\begin{equation}
G(x_f,x_i;k) =  {\mathcal T} \exp[ik (x_f - x_i)],
\label{eq:gdiamond}
\end{equation}
with ${\mathcal T}$ the global transmission amplitude resulting
from the multiple s.p.'s which cross the diamond-shaped region.
By using the procedures in \cite{schmidt2}, one gets
\begin{equation}
{\mathcal T} = \left(\frac{
t_{0 \, +, A} \, P_{+} + t_{0 \, -, A} \, P_{-}}{g}\right) \exp[2 i \gamma].
\label{transmission}
\end{equation}
$P_{+}/g$ ($P_{-}/g$) represents the contribution of all the s.p.'s 
which initially enter the diamond region through the superior 
(inferior) arm. They are given by
($\sigma = \pm$ and $\overline{\sigma} = -\sigma$)
\begin{widetext}
\begin{eqnarray}
P_{\sigma} &=& T_{\sigma \, 0} + 
\Big\{ 
T_{\overline{\sigma} \, 0} \Big( T_{\sigma \, \overline{\sigma}} \, 
r_{\overline{\sigma} \, \overline{\sigma},A} + 
                    R_{\sigma \, \sigma} \, t_{\sigma \overline{\sigma},A} \Big) -
T_{\sigma \, 0}  \Big( T_{\overline{\sigma} \, \sigma} \, 
t_{\sigma \, \overline{\sigma},A} + 
                     R_{\overline{\sigma} \, \overline{\sigma}} 
\, r_{\overline{\sigma} \, \overline{\sigma},A} \Big) 
\Big\}  \exp[2 i \gamma], \nonumber \\
g &=& \Big\{1 -  
\Big( T_{+ \, -} \, t_{- \, +,A} + R_{+ \, +} \, r_{+ \, +,A} \Big) 
\exp[2 i \gamma] \Big\}
 \Big\{1 - 
\Big( T_{- \, +} \, t_{+ \, -,A} + R_{- \, -} \, r_{- \, -,A} \Big) 
\exp[2 i \gamma]  \Big\} 
\nonumber \\
& & - 
\Big( T_{+ \, -} \, r_{- \, -,A} + R_{+ \, +} \, t_{+ \, -,A} \Big)
\Big( T_{- \, +} \, r_{+ \, +,A} + R_{- \, +} \, t_{- \, +,A} \Big)
\exp[4 i \gamma],
\nonumber \\
T_{+ \, 0} &=& 
t_{+ \, -, B} \, \Big\{
t_{+ \, 0,D} +  r_{- \, -,C} \,
\Big(t_{+ \, -,D} \, t_{- \, 0, D} - r_{- \, -,D} \, t_{+ \, 0, D}\Big)
\exp[2 i \gamma] 
\Big\} \exp[i \gamma] \big/f, \nonumber \\
T_{+ \, -} &=& 
t_{+ \, -, B} \, t_{+ \, -, D} \, t_{- \, +, C} \exp[2 i \gamma] \big/f, 
\nonumber \\
R_{+ \, -} &=& r_{+ \, +, B} +  t_{+ \, -, B} 
\Big\{ t_{- \, +, B} \, r_{+ \, +, D} + 
t_{- \, +, B} \, r_{- \, -, C} \,
\Big(t_{- \, +, D} \, t_{+ \, -, D} - r_{+ \, +, D} \, r_{- \, -, D} \Big)
\exp[2 i \gamma] 
\Big\} \exp[2 i \gamma] \big/f,  
\nonumber \\
f &=& \Big(1 -  r_{- \, -, B} \, r_{+ \, +,D} \exp[2 i \gamma] \Big)
\Big(1 - r_{- \, -, C} \, r_{- \, -,D} \exp[2 i \gamma] \Big)
- 
t_{+ \, -, D} \,  t_{- \, +,D} \, r_{- \, -,B} \, r_{- \, -,C}  
\exp[4 i \gamma] 
\nonumber \\
T_{- \, 0} &=& T_{+ \, 0}, \qquad
T_{- \, +} = T_{+ \, -}, \qquad
R_{- \, -} = R_{+ \, +}, \ \mbox{where in all the r.h.s. terms we must exchange} 
\ \
B \longleftrightarrow C.
\label{transmission-coeficients}
\end{eqnarray}
\end{widetext}
For Eqs. (\ref{transmission})-(\ref{transmission-coeficients})  
we have already used the mapping, writing them in terms of  
quantum walk quantities.

The amplitude ${\mathcal T}$ simplifies considerably if 
for any $\sigma$, $\sigma'$, we have $t_{\sigma \, \sigma', X} = t_{X}$ 
and $r_{\sigma \, \sigma, X} = r_{X}$ with $X = A, B, C, D$.
Furthermore, if the coefficients for the sites $A$ and $D$ and for 
the sites $B$ and $C$ are set equal, namely, 
$r_A = r_D$, $t_A = t_D$, $r_B = r_C$ and $t_B = t_C$, Eq.
(\ref{transmission})-(\ref{transmission-coeficients}) yields
\begin{widetext}
\begin{equation}
{\mathcal T} =  \frac{2  \, t_{A}^2 \, t_{B}}
  {1 - 2 \, (t_A + r_A)  \, r_B \exp[2 i \gamma]
  -  (t_A + r_A)^2 \, (t_B^2-r_B^2) \exp[4 i \gamma]}
\exp[3 i \gamma].
\label{transmission-diamond}
\end{equation}
\end{widetext}
For the particular Grover coin values, we get from Eq. 
(\ref{transmission-diamond}) 
${\mathcal T} = 8 \exp[3 i \gamma]/(9-\exp[4 i \gamma])$,
in agreement with Ref. \cite{feldman-pla} as it should be.

We emphasizes that ${\mathcal T}$ is given by a sum over all the 
possible paths leaving $i$, going into the diamond region, and 
finally leaving to the bond $f$.
So, the probability for $i \rightarrow f$ in exactly $n$ 
steps can be obtained by applying the step operator 
to ${\mathcal T}_z = {\mathcal T}|_{\exp[i \gamma] \rightarrow z}$.
Such type of calculation is useful because it gives 
the hitting time $|h_n|^2$ \cite{kempe-hitting}, i.e.,
the probability for the walk to reach a given state 
$|\sigma, j \rangle$ from $|\sigma', j' \rangle$ as function 
of $n$.
The present Green function approach allows to obtain hitting 
times in a rather direct way.
To exemplify this, we first consider the most general case, Eqs. 
(\ref{transmission})-(\ref{transmission-coeficients}), and select 
all the paths reaching the bond $f$ in five time steps. 
Then, $h_5 =  \hat{S}_5 {\mathcal T}_z$, reads
\begin{widetext}
\begin{eqnarray}
h_5 &=&
t_{0 \, +, A} \Big\{ 
            \Big[
                 t_{+ \, -, B} \, r_{+ \, +, D} \, r_{- \, -, B} +
                 r_{+ \, +, B} \, r_{+ \, +, A} \, t_{+ \, -, B}
            \Big] t_{+ \, 0, D} + 
             \Big[
                 t_{+ \, -, B} \, t_{+ \, -, D} \, r_{- \, -, C} +
                 r_{+ \, +, B} \, t_{+ \, -, A} \, t_{+ \, -, C}
            \Big] t_{- \, 0, D} \Big\}
\nonumber \\
& & + t_{0 \, -, A} \Big\{ 
            \Big[
                 t_{+ \, -, C} \, r_{- \, -, D} \, r_{- \, -, C} +
                 r_{+ \, +, C} \, r_{- \, -, A} \, t_{+ \, -, C}
            \Big] t_{- \, 0, D} + 
             \Big[
                 t_{+ \, -, C} \, t_{- \, +, D} \, r_{- \, -, B} +
                 r_{+ \, +, C} \, t_{- \, +, A} \, t_{+ \, -, B}
            \Big] t_{+ \, 0, D} \Big\},
\nonumber \\
\end{eqnarray}
\end{widetext}
which represents the eight possible trajectories with $n=5$,  
Fig. 6 (c).

Certainly, in more symmetric situations analytical results
are easier to obtain.
Indeed, for the case of Eq. (\ref{transmission-diamond}),
${\mathcal T}_z$ can be casted as  
\begin{equation}
{\mathcal T}_z = \frac{-t_A^2}{(t_A + r_A)^2 (t_B^2 - r_B^2)}
\left\{\frac{t_B + r_B}{z^2 - z_-} + \frac{t_B - r_B}{z^2 - z_+}
\right\},
\end{equation}
where $z_{\pm} = (\pm t_B - r_B)/[(t_A + r_A) (t_B^2 - r_B^2)]$.
Hence, for $|h_n|^2 = 
|(n!)^{-1} \, (\partial^n {\mathcal T}_z/\partial z^n)_{z=0}|^2$,
we find that
\begin{widetext}
\begin{eqnarray}
|h_n|^2 &=&
|t_A^2 \, (t_A + r_A)^{(n-1)/2 -1}|^2
\times 
\left\{
    \begin{array}{ll}
    |(t_B + r_B)^{(n-1)/2} - (-1)^{(n-1)/2} (t_B - r_B)^{(n-1)/2}|^2
      & \;\;\;\;\mbox{if} \; n \;\mbox{is odd}, \\
      0
      &\;\;\;\;\mbox{if} \; n \;\mbox{is even},
    \end{array}
\right.
  \label{eq:pt}
\end{eqnarray}
\end{widetext}
is the probability to cross the diamond region in exactly
$n$ steps.
$|h_n|^2 = 0$ for $n < 3$, since at least three time steps are 
necessary to leave the bond $i$ and to arrive at bond $f$. 
Also, from a direct inspection in Fig. 6 one realizes that it 
is not possible a transmission for an even number of steps, a 
result explicit in Eq. (\ref{eq:pt}).
If $t_B = t_C = 0$, obviously the system never gets to the right 
lead and Eqs. (\ref{transmission-diamond}) and (\ref{eq:pt}) are 
identically null.
Finally, if we consider $r_B = 0$ and $t_B = 1$, we find $P_n \neq 0$
only for $n \equiv 3$ (mod 4).
Furthermore, assuming $t_A = 2/3$ and $r_A = -1/3$, we recover the result 
in Ref. \cite{feldman-pla}, namely, 
$|h_n|^2 = (8/9^{(n+1)/4})^2$ for $n \equiv 3$ (mod 4) and 
$|h_n|^2 = 0$ otherwise.

As discussed in the previous section, specific $\mathcal{P}$'s
are obtained from the Green function by means of the
path operator $\hat{P}_{\mathcal P}$.
For instance, suppose we shall select the trajectories 
directly crossing the diamond region, i.e., transmissions 
through $A$, then through $B$ or $C$, and finally through $D$,
with no multiple reflections.
In this special case, the path operator is
\begin{widetext}
\begin{equation}
  \hat{P}_{\mathcal P} = 
  \frac{t_{+ \, -, B}}{1!}\frac{\partial}{\partial t_{+ \, -, B}}
\bigg|_{t = 0 \, (t \neq t_{0 \, +, A}, \, t_{+ \, 0, D}), r = 0} +
 \frac{t_{+ \, -, C}}{1!}\frac{\partial}{\partial t_{+ \, -, C}}
\bigg|_{t = 0 \, (t \neq t_{0 \, -, A}, \, t_{- \, 0, D}), r = 0},
\label{path-example}
\end{equation}
\end{widetext}
leading to
\begin{equation}
\hat{P}_{\mathcal P} \, {\mathcal T} =
  \big(t_{0 \, +, A} \, t_{+ \, -, B} \, t_{+ \, 0, D} +
       t_{0 \, -, A} \, t_{+ \, -, C} \, t_{- \, 0, D}\big)
\exp[3 i \gamma].
\end{equation}
Note that Eq. (\ref{path-example}) is in a simpler form than 
the general definition, Eq. (\ref{path-operator}).
This is so because we have used the fact that $\hat{P}_{\mathcal P}$ 
acts on a transmission Green function.
Indeed, there is no need to perform derivatives as, e.g.,
$t_{0 \, \sigma, A} \, (\partial/\partial t_{0 \, \sigma, A})|_{t_{0 \, \sigma, A}=0}$.
Thus, the path operator is considerably simplified if we
choose suitable configurations to calculate $G$.

We can think of more general paths, namely, to cross 
the diamond region in a total of $n = n_+ + n_-$ 
steps, but for exactly $n_+$ ($n_-$) steps taking in the superior 
(inferior) arm, i.e., in the bonds $A$--$B$ and $B$--$D$ 
($A$--$C$ and $C$--$D$).
If for simplicity we assume that for each site $X$ ($X = B$ or $C$), 
all the $t$'s and $r$'s are equal, regardless the quantum numbers 
$\sigma$'s (as in Eq. (\ref{transmission-diamond})), then the mentioned 
operator, to be applied to ${\mathcal T}$, is written as
\begin{widetext}
\begin{eqnarray}
\hat{P}_{\mathcal P} &=& 
\sum_{n_+^{(1)} +\ldots+ n_+^{(6)} \, = \, n_{+} - 1, \ n_-^{(1)} +\ldots+ n_-^{(6)} \, = \, n_{-} - 1}
       \frac{(t_{- \, +, A})^{n_+^{(1)}}}{n_+^{(1)}!} \,
       \frac{(r_{+ \, +, A})^{n_+^{(2)}}}{n_+^{(2)}!} \,
       \frac{(t_{B})^{n_+^{(3)}}}{n_+^{(3)}!} \,
       \frac{(r_{B})^{n_+^{(4)}}}{n_+^{(4)}!} \,
       \frac{(t_{- \, +, D})^{n_+^{(5)}}}{n_+^{(5)}!} \,
       \frac{(r_{+ \, +, D})^{n_+^{(6)}}}{n_+^{(6)}!} \nonumber \\
& & \times
\frac{(t_{+ \, -, A})^{n_-^{(1)}}}{n_-^{(1)}!} \,
       \frac{(r_{- \, -, A})^{n_-^{(2)}}}{n_-^{(2)}!} \,
       \frac{(t_{C})^{n_-^{(3)}}}{n_-^{(3)}!} \,
       \frac{(r_{C})^{n_-^{(4)}}}{n_-^{(4)}!} \,
       \frac{(t_{+ \, -, D})^{n_-^{(5)}}}{n_-^{(5)}!} \,
       \frac{(r_{- \, -, D})^{n_-^{(6)}}}{n_-^{(6)}!}
\nonumber \\
& & \times \Bigg[
  \Big(\frac{\partial^{n_+^{(1)}}}{\partial t_{- \, +, A}^{n_+^{(1)}}}\Big)
  \Big(\frac{\partial^{n_+^{(2)}}}{\partial r_{+ \, +, A}^{n_+^{(2)}}}\Big)
  \Big(\frac{\partial^{n_+^{(3)}}}{\partial t_{B}^{n_+^{(3)}}}\Big)
  \Big(\frac{\partial^{n_+^{(4)}}}{\partial r_{B}^{n_+^{(4)}}}\Big)
  \Big(\frac{\partial^{n_+^{(5)}}}{\partial t_{- \, +, D}^{n_+^{(5)}}}\Big)
  \Big(\frac{\partial^{n_+^{(6)}}}{\partial r_{+ \, +, D}^{n_+^{(6)}}}\Big)
\nonumber \\
& & \times
\Big(\frac{\partial^{n_-^{(1)}}}{\partial t_{+ \, -, A}^{n_-^{(1)}}}\Big)
  \Big(\frac{\partial^{n_-^{(2)}}}{\partial r_{- \, -, A}^{n_-^{(2)}}}\Big)
  \Big(\frac{\partial^{n_-^{(3)}}}{\partial t_{B}^{n_-^{(3)}}}\Big)
  \Big(\frac{\partial^{n_-^{(4)}}}{\partial r_{B}^{n_-^{(4)}}}\Big)
  \Big(\frac{\partial^{n_-^{(5)}}}{\partial t_{+ \, -, D}^{n_-^{(5)}}}\Big)
  \Big(\frac{\partial^{n_-^{(6)}}}{\partial r_{- \, -, D}^{n_-^{(6)}}}\Big)
  \Bigg]_{t = 0 \, (t \, \neq \, t_{0 \, \pm, A}, \ t_{\pm \, 0, D}), \ r = 0}.
\end{eqnarray}
\end{widetext}

Although the above expression may seem rather cumbersome, it is 
amenable to work with by using algebraic manipulation softwares
(what we have tested by investigating different situations; 
results will be reported elsewhere).
In certain instances, nevertheless, the calculations can be carried
on straightforwardly.
For instance, consider all the paths which get to right lead
only passing through the superior arm.
Furthermore, assume that among them, we shall select those tunneling 
the site $B$ exactly $n$ times.
In this case, the path operator is simply
\begin{widetext}
\begin{equation}
\hat{P}_{\mathcal P} =
\frac{(t_{B})^n}{n!} \, 
\frac{\partial^n}{\partial t_{B}^n}
\bigg|_{t = 0 \, (t \, \neq \, t_{0 \, +, A}, \ t_{+ \, 0, D}), \ r 
= 0 \, (r \, \neq \, r_{+ \, +, A}, \ r_{+ \, +, D})}.
\end{equation}
For $n$ even its results in $\hat{P}_{\mathcal P} {\mathcal T} = 0$,
and for $n$ odd in
\begin{equation}
 \hat{P}_{\mathcal P} {\mathcal T} =
 \frac{t_{0 \, +, A} \, (r_{+ \, +, A})^{\frac{n-1}{2}} 
       \, (t_B)^n \, (r_{+ \, +, D})^{\frac{n-1}{2}}\,  
       t_{+ \, 0,D} \, \exp[(2 n + 1) i \gamma]}
 {(1 - r_{+ \, +, A} r_B \exp[2 i \gamma])^{\frac{n+1}{2}}
  (1 - r_B r_{+ \, +, D} \exp[2 i \gamma])^{\frac{n+1}{2}}}.
\end{equation}
\end{widetext}

\section{Conclusion}

By means of appropriate mappings to systems for which the exact
$G$ can be calculated, quantum graphs, we have obtained closed and general 
expressions for SQW Green functions.
Furthermore, the procedure allows to discuss complete arbitrary
topologies and position dependent quantum amplitudes \cite{energy}.
 
By introducing two operators, namely, step and path operators,
we have shown how to extract from $G$ any relevant dynamical 
information about the system.
For instance, one can exploit particular paths in a quantum walk 
as well as to obtain the contribution of orbits of specific time length
$n$.
Such possibilities have been exemplified in details for a 
diamond-shaped graph structure. 

An important fact, not explored in this contribution, is that
our formulation naturally allows the introduction of energy
(or equivalently, wave number $k$) dependent transition 
amplitudes.
In QW context, such $k$ could be faced as an extra inner variable.
Since different walks may have different $k$'s, using the Green 
function approach, then one could address the case of collective QW.
%Actually, in the literature there are some works treating
%energy dependent CQW \cite{energy}, given raising to different 
%diffusion processes for different choices for the coin energies. 
A complete study of energy dependent SQW will be the subject
of a future work.

Finally, we have discussed $G$ only for QW scattering formulation.
Nevertheless, as already mentioned in the Introduction, the SQW 
and CQW are unitary equivalent.
Moreover, CTQW are associated to CQW.
So, the Green function framework for SQW can be extended to such 
other constructions as well.

\section*{Acknowledgements}
We acknowledge researcher grants by CAPES (F.M.A)
and CNPq (M.G.E.L.).

%\bibliography{green-qw-v1.bib}

\begin{thebibliography}{10}
\bibitem{comment-coef}
Normally, in the context of general point-interactions,
$e_j$ is set to zero.
\bibitem{comment2}
We note there is the possibility of potentials along the
arms (bonds) \cite{kottos1} -- dressed graphs -- but this more general
situation is not relevant for our discussions here.
\end{thebibliography}

\end{document}